\documentstyle[12pt]{article}

\begin{document}

\rightline{TUHE9581}

\begin{center}

{\bf\large The spin of the nucleon in its rest system}\\
\vspace{16pt}

Jerrold Franklin\\
{\it Department of Physics,Temple University,\\
Philadelphia, Pennsylvania 19022}\\
August 1995

\end{center}

\begin{abstract}
The spin dependent structure functions, g$_{1p}$ of the proton and g$_{1n}$ of
the neutron, calculated in the nucleon rest frame using a relativistic quark
model wave function,are compared with recent experiments.
\end{abstract}

\vspace{.5in}

A number of measurements have been made recently of the spin dependent
structure functions, g$_{1p}$ of the proton\cite{emc,smc,e143} and g$_{1n}$ of
the neutron.\cite{e142,smcn}  These measurements have led to some difficulty of
interpretation when used in QCD sum rules, which have seemed to imply a
relatively large polarization of the quark sea (including the strange sea) and
gluon distributions, and a relatively small contribution of quarks to the
proton spin.

An alternative model for calculating nucleon structure has been to calculate
deep inelastic scattering cross sections in the target rest frame, using an
appropriate relativistic quark model wave function.  The structure functions
can then be determined from the cross section.  Details of this model are
contained in Refs. \cite{jf1,ierano,jf2,jf3}.  At its present stage of
development this model does not include quark pair production by the incident
virtual photon, which would correspond to the quark sea that is introduced in
the parton model.  Because of this the fit obtained to the spin independent
stucture functions observed in unpolarized scattering is only good for x$>$0.3,
as is common in valence quark models.  The parameters of the input quark wave
function are determined by a direct fit to the low Q$^2$ (4$<Q^2<20$ GeV$^2$),
large x (x$>$0.3) electron-proton and electron-deuteron SLAC cross section data
summarized by Whitlow.\cite{Whitlow}  This then permits a direct prediction of
the spin dependent structure functions, g$_1$ and g$_2$, of the proton and
neutron.   Somewhat surprisingly, the prediction for g$_1$ agrees with
experiment for x below 0.3 as well as for the higher x.  We take this as an
indication that the mechanisms so far left out of the rest frame model
do not contribute to the polarization.  This is in contrast to the large
polarization of the non-quark effects in the usual parton model interpretation.

We present the comparision with experiment in Figs. $1-5$.  In all cases, the
rest frame structure function has been calculated for the value of Q$^2$
appropriate to each experimental point.  Figure 1 shows the original
EMC\cite{emc} data for g$_{1p}$ along with the rest frame model prediction.
Figure 2 shows the SMC\cite{smc} g$_{1p}$ data with the rest frame model
prediction.   The solid triangle on the x axis represents the x below which
Q$^2$ becomes less than 4 GeV$^2$, and coherent scattering becomes important.
The rest frame model is not expected to apply at these low Q$^2<4$ GeV$^2$.

 Figure 3 shows the g$_{1p}$ data of the SLAC E143 experiment\cite{e143}, along
with the rest frame model prediction.  The points for x$<$0.14, indicated
by the solid triangle on the x axis, have Q$^2<$ 4 GeV$^2$.  At the relatively
low values of Q$^2$ in this experiment, all points for x$>$0.47, indicated by
the open triangle on the x axis, have W (the final state invariant mass) less
than 3 GeV.  The jump in the SLAC g$_{1p}$ in this x region may be due to
direct resonance production polarization, effective at these low values of
W and Q$^2$.  There is only one SMC point from Fig. 2 in this x range,
represented in Fig. 3 by the open circle at x=0.48.  This SMC point has
W=8 GeV and Q$^2$=58 GeV$^2$.  By contrast, the E143 point at
x=0.53 has W=2.8 GeV and Q$^2$=7.6 GeV$^2$. It will be interesting to see if
future data in this x range comes down to the SMC level as Q$^2$ and W are
increased.

The E142 measurement\cite{e142} of the neutron spin dependent structure
function g$_{1n}$ at SLAC is shown in Fig. 4, along with the rest frame model
prediction.  However, all of the E142 data has either W$<$3 GeV or Q$^2<4$
GeV$^2$.  Figure 5 shows the SMC data\cite{smcn} for g$_{1d}$ of the deuteron
normalized to represent the average of g$_{1p}$ and g$_{1n}$, along with the
rest frame model prediction for this.  The points for x below the solid
triangle on the x axis have Q$^2<4$ GeV$^2$.

\end{document}